\documentstyle[multicol,prb,aps,epsf]{revtex}
\renewcommand{\narrowtext}{\begin{multicols}{2}
\global\columnwidth20.5pc\noindent}
\renewcommand{\widetext}{\end{multicols}
\global\columnwidth42.5pc}
\begin{document}

\draft

%
%%%%%%%\twocolumn[\hsize\textwidth\columnwidth\hsize\csname@twocolumnfalse%
%%%%%%%\endcsname
%

\title{Theory of vortex excitation imaging via an NMR relaxation
measurement}

\author{Mitsuaki Takigawa, Masanori Ichioka and Kazushige Machida}
\address{Department of Physics, Okayama University,
         Okayama 700-8530, Japan}
 \date{\today}

\maketitle

\begin{abstract}
The temperature dependence of the site-dependent nuclear spin
relaxation time $T_1$ around vortices is studied in $s$-wave and
$d$-wave superconductors.
Reflecting low energy electronic excitations associated with the
vortex core, temperature dependences deviate from those of the
zero-field case, and $T_1$ becomes faster with approaching the vortex
core.
In the core region, $T_1^{-1}$ has a new peak below $T_c$.
The NMR study by the resonance field dependence may be a new method
to prove the spatial resolved vortex core structure in various
superconductors.
\end{abstract}


%
%]
%

\narrowtext
%%%%%%%%%%%%%%%%%%%%%%%%%%%%%%%%%%%%%%%%%%%%%%%%%%%%%%%

Much attention has been focused on vortex physics both of high $T_c$
cuprates and of conventional superconductors.
Among various experimental methods, the nuclear magnetic resonance
(NMR) experiments\cite{Haase} have been providing vital data in distinguishing
between $s$-wave and $d$-wave pairing symmetries via temperature
($T$) dependence of the nuclear spin relaxation time $T_1$, which
reflects low-lying excitations in the superconducting state.
The power law $T_1^{-1}\propto T^{3}(T^{5})$ behavior is taken as
definitive evidence for a line (point) node in the gap structure of
anisotropic superconductors.
This conclusion comes from a simple counting of the density of states
(DOS) at the Fermi level: $N(\omega)\propto \omega(\omega^{3})$ for a
line (point) node in a bulk superconductor at zero field.
However, actual NMR experiments are performed under applied fields in
a mixed state.
Then, the contribution of the vortex core is included in their
data~\cite{Silbernagel,Ishida}.
Usually, $T_1$ is  measured by selecting the resonance frequency at a
most intensive signal in the resonance spectrum.
However, the resonance spectrum reflects information of internal
magnetic field distribution of the vortex lattice~\cite{Fite}.
By choosing the resonance field, we can specify the position to
detect the NMR signal.
The signal at the maximum (minimum) cutoff comes from the vortex
center (the furthest) site.
The signal at the logarithmic singularity of the resonance field
comes from the saddle points of the field.
By studying the position dependence of $T_1$ around vortices through
the resonance frequency dependence, we can clarify the detail of the
vortex contribution in the NMR experiments.
It helps us in the analysis of the standardized procedure extracting
the gap symmetry.

Low-lying excitation spectra around a vortex are not fully understood
both experimentally and theoretically.
The related problems are as follows.
In the $s$-wave superconductors, the effect of the quantized energy
level will appear in the quasi-particle
state~\cite{CdGM,Caroli,Hayashi}.
In the $d$-wave case, the low energy state around the vortex core
extends outside the core due to the node of the superconducting gap,
and shows the $\sqrt{H}$-like DOS relation ($H$ is an applied
field)~\cite{Volovik,IchiokaDL,IchiokaDS,Wang,Franz}.
We also need to estimate the quasi-particle transfer between vortices
(such as the path of the transfer and its amplitude) to study the
dHvA oscillation or transport phenomena in the mixed
state~\cite{IchiokaDL,IchiokaSL}.
The excitation around the core plays a fundamental role in
determining physical properties of superconductors.
In high $T_c$ cuprates, the existence or non-existence of localized
core excitations in $d$-wave pairing case is actively debated.
Theoretical study suggested the zero-energy peak in the $d$-wave
case, instead of the quantized energy level in the $s$-wave
case~\cite{IchiokaDS,Wang,Franz}.
On the contrary, the scanning tunneling spectroscopy (STS)
experiments reported quantized energy level with large gap in
YBCO~\cite{Maggio}, and surprisingly enough no peak within the
superconducting gap in BSCCO~\cite{Renner}.
A part of reasons of the debate is due to limited experimental
methods which directly probe the spatially resolved core structure.
So far, the STS was only a method to detect it.
A large number of thermodynamic or transport measurements probe
spatially averaged quantities.
Here we propose a novel spatially resolved means, that is, vortex
imaging to see electronic excitations associated with a vortex core
by using NMR, and demonstrate how the $T$-dependence of $T_1$ is
site-sensitive.
Through this analysis, we are able to produce a spatial image of the
low-lying excitation spectrum around a core.
A similar idea of the NMR imaging is actually tested experimentally
in high $T_c$ materials by Slichter's group~\cite{Slichter} and also
in spin-Peiels system CuGeO$_3$ by Horvati\'{c}~\cite{Horvatic}.

The position dependence of the NMR signal in the $s$-wave case was
theoretically studied under some approximations~\cite{Caroli,Leadon}.
Here, we calculate it microscopically from the wave functions
obtained by self-consistently solving the Bogoliubov-de Gennes (BdG)
equation for the extended Hubbard model in the $s$- and $d$-wave
cases.
The eigen-energy $E_\alpha$ and the wave functions $u_\alpha({\bf
r}_i)$, $v_\alpha({\bf r}_i)$ at $i$-site are calculated by following
the method of Ref. \onlinecite{Wang}.
The BdG equation for the extended Hubbard model on the
two-dimensional square lattice is given by
%%%
\begin{equation}
\sum_j
\left( \begin{array}{cc}
K_{i,j} & D_{i,j} \\ D^\ast_{i,j} & -K^\ast_{i,j}
\end{array} \right)
\left( \begin{array}{c} u_\alpha({\bf r}_j) \\ v_\alpha({\bf r}_j)
\end{array}\right)
=E_\alpha
\left( \begin{array}{c} u_\alpha({\bf r}_i) \\ v_\alpha({\bf r}_i)
\end{array}\right) ,
\label{eq:BdG1}
\end{equation}
%%%
where
$K_{i,j}=-t_{i,j} \exp[ {\rm i}\frac{\pi}{\phi_0}\int_{{\bf
r}_i}^{{\bf r}_j} {\bf A}({\bf r}) \cdot d{\bf r} ] - \delta_{i,j}
\mu $, $D_{i,j}=\delta_{i,j} U \Delta_{i,i} + \frac{1}{2}V_{i,j}
\Delta_{i,j}$ with the on-site interaction $U$, the chemical
potential $\mu$ and the flux quantum $\phi_0$.
The transfer integral $t_{i,j}=t$ and the nearest neighber(NN) interaction $V_{i,j}=V$
for the NN site pair ${\bf r}_i$ and ${\bf r}_j$, and otherwise
$t_{i,j}=V_{i,j}=0$.
The vector potential ${\bf A}({\bf r})=\frac{1}{2}{\bf H}\times{\bf
r}$ in the symmetric gauge.
The self-consistent condition for the pair potential is
%%%
\begin{equation}
\Delta_{i,j}=-\frac{1}{2}\sum_\alpha u_\alpha({\bf r}_i)
v^\ast_\alpha({\bf r}_j) \tanh(E_\alpha /2T) .
\label{eq:BdGsc}
\end{equation}
%%%
The band filling factor $\langle n \rangle \sim 0.9$ in our
calculation.

We consider the square vortex lattice case where nearest neighbor
vortex is located at the $45^\circ$ direction from the $a$ axis.
This vortex lattice configuration is suggested for $d$-wave
superconductors, or $s$-wave superconductors with fourfold symmetric
Fermi surface~\cite{IchiokaDL,square-vortex,Won}.
The unit cell in our calculation is the square area of $N_r^2 $ sites
where two vortices are included.
Then, $H=2 \phi_0 /(c N_r)^2$ ($c$ is the atomic lattice constant).
We consider the area of $N_k^2$  unit cells.
By introducing the quasi-momentum of the magnetic Bloch state,
${\bf k}=(2 \pi / c N_r N_k)(l_x,l_y)$ ($l_x,l_y=1,\cdots, N_k$),
we set
$ u_\alpha({\bf r})=\tilde{u}_\alpha({\bf r})
{\rm e}^{{\rm i} {\bf k}\cdot{\bf r}}$,
$v_\alpha({\bf r})=\tilde{v}_\alpha({\bf r})
{\rm e}^{{\rm i} {\bf k}\cdot{\bf r}}$.
We solve Eq. (\ref{eq:BdG1}) within a unit cell under the given
${\bf k}$.
Then, $\alpha$ is labeled by ${\bf k}$ and the eigen-values obtained
by this calculation within a unit cell.

The periodic boundary condition is given by the symmetry for  the
translation ${\bf R}=m{\bf u}_1 +n{\bf u}_2$ ($m$ and $n$ are
integers, ${\bf u}_1$ and ${\bf u}_2$ are unit vectors of the vortex
lattice), i.e., $\tilde{u}_\alpha({\bf r}+{\bf
R})=\tilde{u}_\alpha({\bf r}) {\rm e}^{i\chi({\bf r},{\bf R})/2}$,
$\tilde{v}_\alpha({\bf r}+{\bf R})=\tilde{v}_\alpha({\bf r})
{\rm e}^{-i\chi({\bf r},{\bf R})/2}$.
Here,
$\chi({\bf r},{\bf R})
= -\frac{2\pi}{\phi_0}{\bf A}({\bf r})\cdot{\bf r}
- \pi m n + \frac{2 \pi}{\phi_0}
({\bf H}\times {\bf r}_0)\cdot{\bf R} $
in the symmetric gauge when the vortex center is located at
${\bf r}_0+\frac{1}{2}({\bf u}_1+{\bf u}_2)$.
The on-site $s$-wave pair potential
$\Delta_s({\bf r}_i)=U\Delta_{i,i}$.
The $d_{x^2-y^2}$-wave pair potential is given by
%%%
\begin{equation}
\Delta_{d}({\bf r}_i)=V(\Delta_{\hat{x},i} + \Delta_{-\hat{x},i}
- \Delta_{\hat{y},i} - \Delta_{-\hat{y},i} )/4
\label{eq:dOP1}
\end{equation}
%%%
with
$\Delta_{\pm\hat{e},i}=\Delta_{i,i \pm \hat{e}}
\exp[{\rm i}\frac{\pi}{\phi_0}
\int_{{\bf r}_i}^{({\bf r}_i+{\bf r}_{i \pm \hat{e}})/2}
{\bf A}({\bf r}) \cdot d{\bf r} ]$.
The phase factor~\cite{Ozaki} is needed to satisfy the
translational relation $\Delta_{d}({\bf r})
=\Delta_{d}({\bf r}+{\bf R}){\rm e}^{{\rm i}\chi({\bf r},{\bf R})}$.

We construct the Green's functions from $E_\alpha$, $u_\alpha({\bf
r})$, $v_\alpha({\bf r})$, and calculate the spin-spin correlation
function $\chi_{+,-}({\bf r},{\bf r}',i \Omega_n)$\cite{Leadon}.
Then, we obtain the nuclear spin relaxation rate,
%%%
\begin{eqnarray}
R({\bf r},{\bf r}') &=&
{\rm Im}\chi_{+,-}({\bf r},{\bf r}',
i \Omega_n \rightarrow \Omega + {\rm i}\eta)/(\Omega/T)|_{\Omega
\rightarrow 0}
\nonumber \\
&=&
 -\sum_{\alpha,\alpha'} u_\alpha({\bf r})u^\ast_{\alpha'}({\bf r})
[ u_\alpha({\bf r}')u^\ast_{\alpha'}({\bf r}')
 +v_\alpha({\bf r}')v^\ast_{\alpha'}({\bf r}') ]
\nonumber \\ &&
\times \pi T f'(E_\alpha) \delta(E_\alpha - E_{\alpha'})
\label{eq:T1}
\end{eqnarray}
%%%
with the Fermi distribution function $f(E)$.
We consider the case ${\bf r}={\bf r}'$ by assuming that
the nuclear relaxation occurs locally such as in Cu-site
of high $T_c$ cuprates.
%It is noted that the relaxation occurs within Cu-site in high-$T_c$
% cuprates.
Then, ${\bf r}$-dependent relaxation time is given by $T_1({\bf
r})=1/R({\bf r},{\bf r})$.
In Eq. (\ref{eq:T1}), we use $\delta(x)=\pi^{-1} {\rm Im}(x-{\rm
i}\eta)^{-1}$ to consider the discrete energy level of the finite
size calculation.
We typically use $\eta=0.01t$.
To understand the behavior of $T_1({\bf r})$, we also consider the
local density of states (LDOS) given by
%%%
\begin{eqnarray} &&
N({\bf r},E)
\nonumber \\ &&
=-\sum_\alpha [|u_\alpha ({\bf r})|^2f'(E_\alpha -E)
+ |v_\alpha ({\bf r})|^2f'(E_\alpha +E)].
\label{eq:STM}
\end{eqnarray}
%%%
It corresponds to the differential tunnel conductance of STS
experiments.

% ###### FIG. 1 start ###################################################
%%%%%%%%%%%%%%%%%%%%%%%%%\end{twocolumn}
\widetext
\begin{figure}[tbp]
\epsfxsize=16cm
\hspace{5.5mm}
\epsfbox{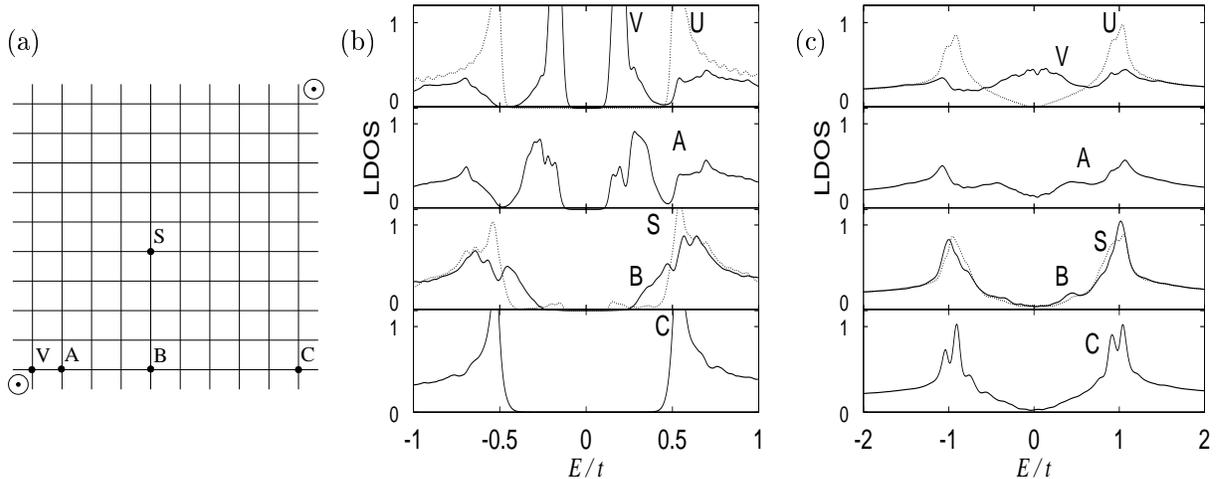}
\vspace{5.5mm}
\caption{
The LDOS $N({\bf r},E)$ at sites V, A, B, C, S.
(a) Position of the sites V, A, B, C, S in the square vortex lattice,
where the nearest neighbor vortex is located in the $45^\circ$
direction from the $a$-axis.
The vortex center is shown by $\odot$.
The solid lines show the square atomic lattice.
(b) $N({\bf r},E)$ in the $s$-wave case.
(c) $N({\bf r},E)$ in the $d$-wave case.
The LDOS for the S-site is presented by the dotted line with the
solid line for the B-site.
The DOS at zero-field is presented by the dotted line U with the
solid line for the V-site.
}
\label{fig:LDOS}
\end{figure}
\narrowtext
% ###### FIG. 1  end ####################################################

As for the temperature dependence of $\Delta_s({\bf r})$ and
$\Delta_d({\bf r})$, the vortex core radius shrinks with decreasing
$T$ by the Kramer-Pesch effect~\cite{IchiokaDS,Kramer}.
However the shrink is saturated at a low temperature both in the $s$-
and $d$-wave cases.
There, the structure of $\Delta_s({\bf r})$ and $\Delta_d({\bf r})$
is almost independent of $T$.
This is a quantum-limit effect which occurs for $T/T_c <
\Delta_0/E_{\rm F}$ ($E_{\rm F}$ is the Fermi energy and $\Delta_0$
the superconducting gap at zero field. In our calculation,
$\Delta_0/E_{\rm F} \sim 0.25$ for $d$-wave, $\sim 0.125$ for $s$-wave)
~\cite{Hayashi}.
We calculate the low temperature behavior of $T_1({\bf r})$ by using
the saturated pair potential.
At higher temperature, we calculate $T_1({\bf r})$ by using the
self-consistently obtained pair potential at each $T$.

Figure \ref{fig:LDOS} (a) shows the position of the sites
 (V, A, B, C, S) where we calculate $N(E,{\bf r})$
and $T_1({\bf r})$.
First, we see the LDOS around the vortex.
The $s$-wave case ($U= - 2.32t$, $V=0$) is shown in Fig.
\ref{fig:LDOS}(b), and the $d$-wave case ($U=0$, $V=-4.20t$) is shown in
Fig. \ref{fig:LDOS}(c).
In our calculation, $N_r=20$ and $N_k=8$.
In $N(E,{\bf r})$ at the vortex center (the V-site), the gap edge at
$\Delta_0$ in the zero-field case (dotted line U)  is smeared, and
low-energy peaks of the vortex core state appear.
In the $s$-wave case, we see some peaks above the small gap
$\Delta_1$ ($\sim \Delta_0^2/E_{\rm F})$.
It is due to the quantization of the energy level in the $s$-wave
case.
In the $d$-wave case, the core state shows zero-energy peak instead
of the split peaks in the $s$-wave case~\cite{Wang}.
There is no small gap.
The weight of the low-energy states is decreased with going away from
the vortex center (V$\rightarrow$A$\rightarrow$B$\rightarrow$C).
Far from the vortex, $N(E,{\bf r})$ is reduced to the DOS of the
zero-field case.
But, small weight of the low-energy state extending from the vortex
core remains there.
It is noted that the weight of the low-energy state at the S-site is
larger than that of the B-site in the $s$-wave case, while the S-site
is farther from the vortex center [see lines for the S- and B-sites
in Fig \ref{fig:LDOS}(b)].
It is due to the vortex lattice effect.
The quasiparticle transfer between vortices occurs along the line
connecting NN vortices (i.e., near the S-site).

Next, we consider the $T$-dependence of $T_1({\bf r})$ at each site
in Fig. \ref{fig:LDOS}, which reflects the LDOS discussed above.
The NMR signal at the maximum cutoff of the resonance spectrum
as a function of applied field or probe frequency
comes from the vortex core at the V-site.
With going away from the center
(V$\rightarrow$A$\rightarrow$B$\rightarrow$C), the resonance field is
decreased.
The signal at the minimum cutoff comes from the C-site.
The logarithmic singularity of the resonance field comes from the
saddle point of the field at the S-site.
Thus it is possible to perform the site-selective $T_1({\bf r})$
measument by tuning the resonance frequency.

\begin{figure}[tbp]
\epsfxsize=85mm
%\hspace{5.5mm}
\epsfbox{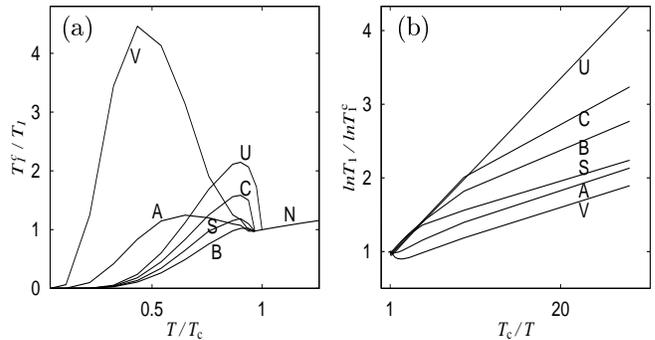}
%\vspace{5.5mm}
\vspace{6.0mm}
\caption{
Temperature dependence of $T_1({\bf r})$ in the $s$-wave case at the
sites V, A, B, C, S assigned in Fig. \protect\ref{fig:LDOS}(a).
(a) $T_1(T_c)/T_1(T)$ is plotted as a function of $T/T_c$.
(b) $\ln T_1(T)/\ln T_1(T_c)$ is plotted as a function of $T_c/T$.
Line U shows the zero field case.
The line N is for the normal state at $T>T_c$.
}
\label{fig:T1s}
\end{figure}

The $s$-wave case is shown in Fig. \ref{fig:T1s}.
We plot $T_1({\bf r})^{-1}$ vs. $T$ for each site in Fig.
\ref{fig:T1s}(a),  and re-plot it as $\ln T_1({\bf r})$ vs. $T^{-1}$
in Fig. \ref{fig:T1s}(b).
We also calculate the zero-field case in our formulation.
At the zero field, $T_1 \sim {\rm e}^{\Delta_0 /T}$.
Then, the slope of the $\ln T_1$ vs. $T^{-1}$ plot gives the
superconducting gap $\Delta_0$, as the line U in Fig.
\ref{fig:T1s}(b).
In the presence of vortices, $T_1$ deviates from the relation ${\rm
e}^{\Delta_0 /T}$ at low $T$ due to the low-energy excitation around
the vortex core.
This deviation was reported in the experiments~\cite{Silbernagel}.
In our results, reflecting the small gap $\Delta_1$ in the $s$-wave
case, $T_1$ shows the slope $\Delta_1$ at low $T$ in the $\ln T_1$
vs. $T^{-1}$ plot(see the V-site in Fig.1(b))
 as seen in Fig. \ref{fig:T1s}(b).
That is, $T_1 \sim {\rm e}^{\Delta_1 /T}$.
With leaving the vortex center, since the amplitude of the low-energy
bound states is damped, the weight of ${\rm e}^{\Delta_1 /T}$
gradually decreases.
Then the crossover temperature from ${\rm e}^{\Delta_0 /T}$ to
${\rm e}^{\Delta_1 /T}$ is lowered.
It is noted that $T_1$ is faster at the S-site than that of the
B-site, while the S-site is further from the vortex center.
This non-trivial result is due to the vortex lattice effect noted above.
We should also notice the behavior of the coherence peak below
$T_c$.
As seen in Fig. \ref{fig:T1s}(a), with approaching the vortex center
as C$\rightarrow$B, the coherence peak is suppressed.
But in the vortex core region (lines V and A), a large new peak
grows at intermediate temperatures.
This is because the LDOS at the vortex core has peaks at low energy
$\Delta_1$ instead of the singularity of DOS at $\Delta_0$.

% ###### FIG. 2 start ###################################################
\vspace{5.5mm}
% ###### FIG. 2  end ####################################################

% ###### FIG. 3 start ###################################################
\begin{figure}[tbp]
\epsfxsize=85mm
%\hspace{5.5mm}
\epsfbox{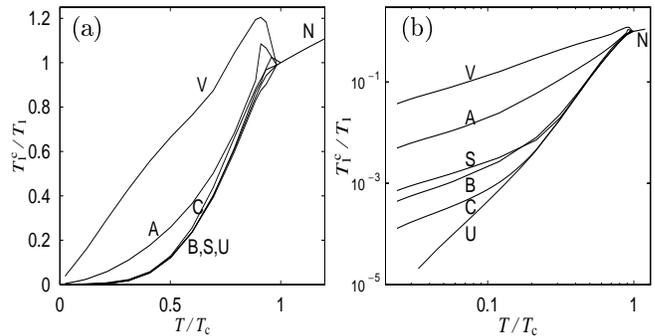}
%\vspace{5.5mm}
\vspace{6.0mm}
\caption{
Temperature dependence of $T_1({\bf r})$ in the $d$-wave case at the
sites V, A, B, C, S.
(a) $T_1(T_c)/T_1(T)$ is plotted as a function of $T/T_c$.
(b) a log-log plot of (a).
Line U shows the zero field case.
The line N is for the normal state at $T>T_c$.
}
\label{fig:T1d}
\end{figure}
% ###### FIG. 3  end ####################################################

As for the $d$-wave case, we plot $T_1({\bf r})^{-1}$ vs. $T$ in Fig.
\ref{fig:T1d}(a), and re-plot it as a log-log plot in Fig.
\ref{fig:T1d}(b).
At zero field (line U), we see the power law relation $T_1^{-1} \sim
T^3$ of the $d$-wave case as expected.
Note that this can be seen only below $T/T_c \simeq 0.1$ in our case.
In the presence of vortices, $T_1({\bf r})^{-1}$ deviates from the
$T^3$-relation, and follows $T_1({\bf r})^{-1} \sim T$ at low
temperature.
This deviation was reported in the experiments on high-$T_c$
cuprates~\cite{Ishida}.
The origin of the $T$-linear behavior, which is attributed
to residual density of states due to
impurities or defects, is the low-energy state around vortices in our case.
With approaching the vortex center, the $T$ region of the
$T$-linear behavior is enlarged and it appears from higher
temperatures.
As seen in Fig. \ref{fig:LDOS}(c) of the $d$-wave case, the
superconducting gap is buried by the low-energy state around vortices
without the small gap of the order $\Delta_0^2/E_{\rm F}$.
Then, $T_1^{-1} \sim T$ at low temperature in the $d$-wave case
instead of the relation $T_1 \sim {\rm e}^{\Delta_1 /T}$ in the
$s$-wave case.
As seen in Fig. \ref{fig:T1d}, $T_1({\bf r})^{-1}$ at the vortex
center  (line V) is very large compared with the zero-field case
(line U).
It reflects the fact that the LDOS of the low-energy state is larger
than the DOS of the zero-field case as seen in Fig. \ref{fig:LDOS}(c).
This short relaxation may be the evidence of the low-energy peak in
the LDOS by the low-energy core state.
The coherence peak below $T_c$ is taken as a manifestation of
the $s$-wave symmetry.
In the $d$-wave case, the coherence peak is absent.
But in the vortex core region, $T_1^{-1}$ has a peak below $T_c$ even
in the $d$-wave case.
We should be careful not to mistake this peak due to the vortex core
relaxation as the usual coherence peak in the NMR experiment when
identifying the gap symmetry.

With increasing external magnetic field, the relaxation is enhanced,
because the vortex contribution is increased and the amplitude of the
low energy state extending outside the vortex core becomes large
both in the $s$-wave and $d$-wave cases, as coinciding qualitatively
with the observation of an orgnanic superconductor
$\kappa$-(ET)$_2$Cu[N(CN)$_2$]Br by Mayaffre {\it et al.}~\cite{Mayaffre}.
The details of the field dependence belong to a future study.

Traditionally, the vortex contribution was considered as the spin
diffusion to the normal region of the vortex
core~\cite{Ishida,Caroli}, and $T_1$ is treated as the spatial average.
However, we can investigate the position dependence of $T_1({\bf r})$
around vortices through the resonance field dependence.
This is an advantage of NMR over other methods.
We should clarify the local mechanism of the relaxation (i.e.,
whether the relaxation occurs locally, or it is averaged by the spin
diffusion).
It is noted that in the clean limit the vortex core region is
not a simple core filled by normal state electrons~\cite{IchiokaDL}.
There, the characteristic $T$-dependence is expected near the vortex
core  other than a simple $T$-linear behavior, reflecting the rich
structure of the low energy state around the vortex core.
We expect that the NMR imaging study  just explained here
will provide vital information for the vortex core state
in high-$T_c$ cuprates.
As for the problem whether the quantization of the energy levels
occurs or not, $T_1 \sim {\rm e}^{\Delta_1/T}$ if the gap $\Delta_1$
($\sim \Delta_0^2/E_{\rm F}$) is present in the excitation due to the
quantization.
If this small gap is absent, $T_1^{-1} \sim T$.
As for the problem whether the zero-energy peak exists or not in the
core state, the relaxation at the core becomes eminently faster than
that of the zero-field case (or that far from the vortex) at low
temperature, if the zero-energy peak exists in the LDOS as suggested
in the theoretical study.
If the peak structure is absent within $\Delta_0$ as reported in the
STM experiments on BSCCO, the relaxation is slow even at the vortex
core as in the zero-field superconducting case.

We proposed the study of the low-energy excitation imaging around
vortices via an NMR  relaxation.
It may provide valuable information for the understanding of the
vortex physics  in high-$T_c$ superconductors as well as in the
conventional superconductors.

We thank M. Horvati\'{c}, K. Ishida and Y. Iwamoto
for useful infomration on NMR experiments.

%%%% references %%%%%%%%%%%%%%%%%%%%%%%%%%%%%%%%%%%%%%%

%%%%%%%%%%%%%%%%%%%%
%\newpage
%%%%%%%%%%%%%%%%%%%%

\widetext

\end{document}